\magnification = \magstep1
\vsize = 22 true cm
\hsize = 16 true cm
\baselineskip = 24 true pt
\centerline {\bf {CONFORMALLY INVARIANT PATH INTEGRAL FORMULATION OF}} 
\centerline {\bf {THE WESS-ZUMINO-WITTEN $\rightarrow$ LIOUVILLE REDUCTION}} 
\bigskip
\bigskip
\centerline {L. O'Raifeartaigh and V. V. Sreedhar} 
\centerline {School of Theoretical Physics}
\centerline {Dublin Institute for Advanced Studies} 
\centerline {10, Burlington Road, Dublin 4}
\centerline {Ireland}
\bigskip
\bigskip
\centerline {\bf {Abstract}}
The path integral description of the Wess-Zumino-Witten $\rightarrow$ Liouville
reduction is formulated in a manner that exhibits the conformal invariance 
explicitly at each stage of the reduction process. The description  
requires a  conformally invariant generalization of the phase space path 
integral methods of  Batalin, Fradkin, and Vilkovisky for systems with 
first class constraints. The conformal anomaly is incorporated 
in a natural way and a generalization 
of the Fradkin-Vilkovisky theorem 
regarding gauge independence is proved. 
This generalised formalism should apply to all conformally invariant 
reductions in all dimensions. A previous problem concerning the gauge  
dependence of the centre of the Virasoro algebra of the reduced theory 
is solved.  
\vfill
\hfill DIAS-STP-97-13\hfil\break
\vfil\eject
\centerline {\bf {I.  INTRODUCTION AND THE STATEMENT OF THE PROBLEM}}
\bigskip
In the course of the past decade the classical Hamiltonian reduction of 
Wess-Zumino-Witten (WZW) theories to Toda theories using first class 
constraints, and the concomitant reduction of Kac-Moody algebras to W-algebras, 
has been formulated in considerable detail [1]. The quantized version of the 
reduction process has also been considered, but  mainly within the framework of 
canonical quantization [2]. The elegance of the classical reduction process 
suggests, however, that the most natural framework for quantization  
is through the functional integral. Accordingly, 
in this paper, we present the functional integral formulation for the 
quantization of the simplest WZW $\rightarrow$ Toda reduction, namely the 
reduction of the $SL (2, R)$ WZW theory to the Liouville theory. More general   
cases may be dealt with in an analogous fashion and will be considered 
later. It turns out that a suitable refinement of the 
Batalin-Fradkin-Vilkovisky 
(BFV) formalism for constrained systems [3] 
introduced in this paper, does indeed allow the functional integral reduction
to proceed in an elegant manner. 

The setting up of the functional integral reduction process presents a few  
subtleties that make it worthwhile to present our results in some  
detail. The main point is that the WZW $\rightarrow$ Liouville reduction  
should be conformally invariant but neither the 
usual Faddeev-Popov-BRST formalism, nor its BFV generalization 
guarantees this. These formulations  are primarily  
concerned with gauge invariance and to make them conformally invariant 
as well requires a non-trivial extension, especially in view of the 
conformal anomaly.  
We find such  
an extension, and within this generalised formalism, prove a conformally  
invariant generalization of the Fradkin-Vilkovisky theorem regarding
gauge 
independence. An important point, 
although we have not pursued it in this paper, is that the same procedure 
should be valid for any conformally invariant reduction and for any number of 
dimensions.

The failure of the straighforward Faddeev-Popov-BRST method was pointed out 
in an earlier paper [4], where it took the form of a discrepancy between the 
values of the Virasoro centres obtained in two different gauges. More 
precisely, it turned out that while the two centres had the same functional 
form, the arguments were $k$ and $k - 2$ where $k$ is proportional to the 
WZW coupling constant $\kappa$. Such a discrepancy suggests, of course, the 
existence of a conformal anomaly. But a straightforward application of the  
Faddeev-Popov method, or even the usual BFV method,  produces no such anomaly. 
The generalised formalism that is developed in this paper, which allows   
us to keep track of both gauge and conformal invariances  at   
each stage of the reduction, resolves this problem.    

There are a number of other novel situations that arise in the 
WZW $\rightarrow$ Liouville reduction. First, we note that although, 
as usual,  the original WZW  
Hamiltonian is not bounded below because it is based on a non-compact group, 
the Hamiltonian for the reduced theory is  
positive definite and is thus physically acceptable.   
More importantly, the fact that it is not possible 
to choose configurations such that both the kinetic term and the 
potential of the Liouville Action are simultaneously finite on a 
non-compact base-space 
means that the base-space must be compact. As a consequence of this, one  
has to beware of zero modes when gauge fixing.    
In fact it is the gauge-invariant zero-modes which actually produce the 
Liouville interaction term in the reduced theory.  

This paper is organised in the following manner. In section two we review 
the Hamiltonian formalism of the $SL(2, R)$ WZW model and sketch the
essential ingredients of the classical reduction procedure.  
As the BFV formalism is the natural one to use for the path integral approach, 
the basic structure of this formalism is  presented  
in section three. The heart of the paper is contained in section four in
which we formulate the conformally invariant generalization of the BFV  
formalism, and establish its  
gauge invariance by proving the analogue of the Fradkin-Vilkovisky theorem.  
In section five we compare the path integral reduction process
as formulated in this paper with earlier attempts in this direction. In the  
sixth and final section, we give a summary of our results. 
\bigskip
\centerline {\bf{II. THE CLASSICAL SL(2, R) WZW $\rightarrow$ LIOUVILLE 
REDUCTION}}
\bigskip
The WZW model is defined on a two dimensional manifold $\partial\Sigma$ 
by the Action [5]    
$$S_{WZW} = {\kappa\over 2\pi}\int_{\partial\Sigma} Tr~(g^{-1}d g)
\cdot (g^{-1}d g) - {\kappa\over 3\pi}\int_\Sigma Tr~ 
(g^{-1}d g)\wedge (g^{-1} d g)\wedge (g^{-1}d g)
\eqno(2.1) $$
In the above $g\in G\equiv SL(2, R)$. In what follows we shall set the
coupling constant ${\kappa\over \pi}$ equal to one, for convenience, and 
restore it when it becomes of interest in section four. 
The two dimensional manifold is parametrized by the light-cone coordinates 
$z_r$ and $z_l$ defined by 
$$z_r = {z_0 + z_1 \over 2},~~~z_l = {z_0 - z_1\over 2}\eqno(2.2)$$
The Action is invariant under 
$$g\rightarrow gu(z_r ),~~~ g\rightarrow v(z_l )g \eqno(2.3)$$
where $u(z_r), v(z_l) \in G$.  
The conserved Noether currents which generate the above transformations are 
given by 
$$J_r = (\partial_r g)g^{-1}, ~~~J_l = g^{-1}(\partial_l g)\eqno(2.4)$$
and take their values in the infinite dimensional Lie algebra of the model. 
In order to set up the Hamiltonian formalism, let us introduce the Gauss 
decomposition for the group-valued field $g$
$$g = \exp {(\alpha\sigma_+)} \exp {(\beta\sigma_3)} \exp {(\gamma\sigma_-)}
\eqno(2.5)$$ 
where $\sigma_\pm$ and $\sigma_3$ are the generators of the $SL(2, R)$ Lie 
algebra. 
$$\sigma_+ = {\pmatrix {0&1\cr 0&0\cr}}, ~~~\sigma_- = 
{\pmatrix {0&0\cr 1&0\cr }},~~~\sigma_3 = {\pmatrix{1&0\cr 0& -1}}\eqno(2.6)$$ 
As is well-known, the Gauss decomposition is not valid globally. This issue 
has been dealt with in detail in [6]. For simplicity, we restrict  
our present considerations to the coordinate patch that contains the 
identity. Similar results hold for the other patches.  
In terms of the local coordinates $\alpha , \beta ,\gamma $ on the group 
manifold the Action can be rewritten as 
$$S_{WZW} = \int d^2z~ \left[ (\partial_\mu\beta )(\partial^\mu\beta ) + 
(\partial_l\alpha )(\partial_r\gamma )\exp (-2\beta )\right] \eqno (2.7)$$
The momenta canonically conjugate to $\alpha , \beta , \gamma $ respectively 
are defined, as usual, by 
$$\pi_\alpha = {\delta{\cal {L}}\over \delta (\partial_0\alpha )} = 
(\partial_r\gamma ){\exp (-2\beta )} \eqno(2.8a)$$
$$\pi_\gamma = {\delta{\cal {L}}\over \delta (\partial_0\gamma )} = 
(\partial_l\alpha ){\exp (-2\beta )} \eqno(2.8b)$$
$$\pi_\beta = {\delta{\cal {L}}\over \delta (\partial_0\beta )} = 
2\partial_0\beta \eqno(2.8c)$$ 
The canonical Hamiltonian density $H_{WZ}$ is 
$$ H_{WZW} = {1\over 4}\pi_\beta^2 +  (\beta ' )^2 + 
\pi_\alpha\pi_\gamma\exp (2\beta)+ \pi_\alpha
\alpha ' - \pi_\gamma \gamma ' \eqno(2.9)$$
The currents can be expanded in the basis of the Lie algebra and the 
various components can be read off from the following equations 
$${\pmatrix {J_r^+ \cr J_r^3 \cr J_r^- }} = 
{\pmatrix { 1 & -2\alpha & -\alpha^2\exp (-2\beta ) \cr 
0 & 1 & \alpha\exp (-2\beta )\cr 0 & 0 & \exp (-2\beta )}}{\pmatrix 
{\partial_r\alpha\cr \partial_r\beta\cr \partial_r\gamma} }\eqno(2.10a)  $$
$${\pmatrix {J_l^+ \cr J_l^3 \cr J_l^- }} = 
{\pmatrix { \exp (-2\beta ) & 0&0\cr \gamma\exp (-2\beta )&
1&0\cr 
-\gamma^2\exp (-2\beta )& -2\gamma & 1}}
{\pmatrix {\partial_l\alpha \cr
\partial_l\beta \cr \partial_l\gamma }} \eqno(2.10b)$$ 
The currents may also be expressed completely in terms of the phase space 
variables $\alpha ,\beta ,\gamma$ and their conjugate momenta using the  
relations in Eq.(2.8).  Further, by using canonical Poisson brackets  
for the phase space variables {\it viz.}
$$\{\alpha (z) ,\pi_\alpha (z')\} = \{\beta (z),\pi_\beta (z')\} = 
\{\gamma (z) ,\pi_\gamma (z')\} = \delta (z-z')\eqno(2.11)$$
the rest being zero, we can check explicitly that the currents satisfy two 
independent copies of the infinite dimensional Kac-Moody algebra 
$$\eqalign {\{ J_r^3 (z_r), J_r^{\pm} (z_r')&\} = \pm J_r^{\pm }
\delta (z_r - z_r'), ~~~~  
\{J_r^3 (z_r), J_r^3(z_r')\}
= \partial_{z_r}\delta (z_r-z_r') \cr 
&\{J_r^+(z_r), J_r^-(z_r')\} = 2(J_r^3 - \partial_{z_r'})\delta (z_r-z_r')}
\eqno(2.12)$$
Similar equations are valid for the left currents.  
In terms of the currents, the Hamiltonian density $ H$ can be written in
the Sugawara form 
$$  H_{WZW} = {1\over 2}\{J_r^+J_r^- + (J_r^3)^2 + J_l^+J_l^- +
 (J_l^3)^2\} \eqno(2.13)$$
The constraints we want to impose are 
$$ \phi_r \equiv J_r^- - m_r  \approx 0, ~~~\phi_l\equiv J_l^+ - 
m_l  \approx 0 \eqno(2.14a)$$
or equivalently, 
$$ \phi_r \equiv \pi_\alpha - m_r \approx 0,~~~\phi_l\equiv \pi_\gamma 
-m_l \approx 0 \eqno(2.14b)$$  
where $m_r $ and $m_l $ are constants. 
However, these constraints are not consistent with the conformal invariance 
defined by the two Sugawara Virasoro operators  
$${\cal T}_r =  {1\over 2}
\{J_r^+J_r^- + (J_r^3)^2 \}, ~~~
 {\cal T}_l = {1\over 2}\{ J_l^+J_l^- + (J_l^3)^2\}\eqno(2.15)$$ 
because, as is well-known, the currents $J_r^-$ and $J_l^+$ are not conformal 
scalars, but spin one fields.   
However, taking advantage of the fact that the Virasoros 
for Kac-Moody algebras are unique only up to the addition of a diagonalizable 
element of the algebra or its first derivative, we modify the
Sugawara Virasoros  
above to define the components of the so-called improved energy momentum  
tensor, namely $T_r$ and $T_l$.  
$$T_r  = {1\over 2}\{J_r^+ J_r^- + (J_r^3)^2 -2\partial_r J_r^3\}\eqno(2.16a)$$
$$T_l = {1\over 2}\{J_l^+J_l^- + (J_l^3)^2 +2 \partial_l J_l^3\}\eqno(2.16b)$$
The physical meaning of the additional terms is that they are just the 
`improvement'  terms necessary to make the energy-momentum tensor of the  
reduced theory traceless. From Eqs.(2.13) and (2.16), it is clear  
that the above modification is tantamount to adding only total derivative 
terms to the Hamiltonian density. 
Hence this modification leaves the Hamiltonian, and consequently the dynamics 
of the theory, invariant. With respect to the conformal group generated 
by the Virasoros (2.16), the currents 
$J_r^-$ and $J_l^+$ are conformal scalars {\it i.e.} they now have   
conformal weights, denoted by $\omega$, as follows   
$$ \omega (J_r^-) = \omega (J_l^+) = (0, 0)\eqno(2.17)$$ 
The constraints in Eq.(2.14) are, therefore, compatible with this   
conformal group.  

The currents $J_r^+$ and 
$J_l^-$ now have conformal weights (0, 2) and (2, 0) respectively.  
The phase space variables $\alpha $ and $\gamma $ become primary fields 
of conformal weights (0, 1) and (1, 0) respectively, the field $\beta $
becomes a conformal connection, while $e^{2\beta} $ becomes a  primary 
field of weight (1, 1) {\it i.e.} it has the opposite conformal weight 
to the volume element $d^2z$ in the two dimensional space.  

Upon imposing the constraints (2.14b) on the classical Hamiltonian density 
(2.9) of the $SL(2, R)$ WZW model, we get, apart from boundary terms, 
$$ H_{reduced} = {1\over 4}\pi_\beta^2 + (\beta ')^2 + m_r 
m_l e^{2\beta } \eqno(2.18)$$ 
This is easily recognised as the expression for the Hamiltonian density of the  
classical Liouville theory. Since the constraints we impose are linear in the  
momenta, it is natural to use the phase space path integral, rather than the 
configuration space path integral, for setting up the functional integral  
formulation of the above classical reduction.  The next section, therefore,   
prepares us for the quantization of this reduction using phase space path 
integral methods. 
\vfil\eject
\centerline {\bf {III. THE BATALIN-FRADKIN-VILKOVISKY PATH INTEGRAL}}
\bigskip
As mentioned in Section I, our aim is to establish a functional integral 
formulation for the WZW $\rightarrow$ Liouville reduction. But since one 
of the gauges we are interested in is the WZW gauge in which the Lagrange 
multipliers are set equal to zero (the analogue of the temporal gauge in 
QED), the standard Faddeev-Popov [FP] method 
does not quite suffice. A more general method for quantizing constrained 
systems, namely the BFV procedure, needs to be used. 
Hence we begin by recalling the basics of the BFV procedure. Let  
$$Z = \int d(pq)~ e^{-\int dxdt~[p\dot q - H(p, q)]}\eqno(3.1)$$ 
where $p$ and $q$ are any set of canonically conjugate variables,    
be the phase space path integral which is to be reduced by a set of 
first class constraints $\Phi (q, p)$. Let $A$ be a set of Lagrange 
multipliers, $B$ their canonically conjugate momenta, and $b, \bar c$ 
and $c, \bar b$ be conjugate ghost pairs. Then define the BRST charge by  
$$\Omega = \int dx~ [c\Phi + bB] + \cdots \eqno(3.2a)$$
where the dots refer to terms which involve higher order ghosts which occur 
in the general case but do not occur in the WZW $\rightarrow$ Liouville 
reduction. The BRST charge $\Omega $ also satisfies the nilpotency   
condition 
$$\{\Omega , \Omega \} = 0 \eqno(3.2b)$$
A minimal gauge fixing fermion $\bar\Psi$ is then defined as  
$$\bar\Psi = \bar c\chi + \bar b A \eqno(3.3)$$    
where $\chi (p, q, A, B) $ is a set of gauge-fixing conditions.  
The BFV procedure now consists of inserting the following reduction factor 
$$ F = \int d(ABb\bar b c\bar c )e^{-\int dxdt~[\bar b \dot c + \{\Omega , 
\bar \Psi \}]}\eqno(3.4)$$ 
into the path integral in Eq. (3.1). 

>From the non-zero Poisson brackets for the variables  
$$ \{q (x), p (x')\} = \{A (x), B(x')\} = 
 \{b (x), \bar c (x')\} = 
 \{c (x), \bar b (x')\} = \delta (x - x') \eqno(3.5)$$  
we see that the gauge variations of the fields are  
$$ \{\Omega , f(q, p)\} = c \{\Phi , f(q, p)\}\eqno(3.6a)$$
$$ \{\Omega , A\} = -b,~~~ \{\Omega , B\} = 0 \eqno(3.6b)$$
$$\{\Omega , \bar b\} = \Phi ,~~~ \{\Omega , \bar c\} = B\eqno(3.6c)$$
$$ \{\Omega , b\} = \{\Omega , c\} = 0 \eqno(3.6d)$$
where $f(q, p)$ is an arbitrary function of the phase space variables.   
It follows from the above equations that 
$$\{\Omega , \bar\Psi\} = (A\Phi + B\chi) + 
(- \bar b b + \bar c [FP]c + \bar c [BFV]b)\eqno(3.7)$$ 
where the FP and BFV terms are defined by 
$$ \{\Phi (x), \chi (x')\} = [FP]\delta (x - x'), ~~
\{B(x), \chi (x')\} = [BFV]\delta (x - x') \eqno(3.8)$$ 
Note that in the definition of the reduction factor above, it is not necessary 
to include the term $B\dot A + {\dot {\bar c}} b$ in the Action. This is
because such a term can always be generated by letting  
$\chi \rightarrow \chi + \bar c\dot A$. By virtue of the 
Fradkin-Vilkovisky theorem, which says that the reduced 
functional integral $Z_R$ 
is independent of the choice of the gauge fixing fermion $\bar\Psi$, 
the above definition of the functional integral produces the correct quantum 
theory. Substituting for $\{\Omega , \bar\Psi\}$ in $F$ and doing the 
$\bar b b$ integrations yields 
$$ F = \int d(AB\bar cc) e^{-\int dxdt ~[A\Phi + B\chi + \bar c\{[FP] + 
[BFV]\partial_t \}c ] }\eqno(3.9)$$
Inserting this factor into Eq. (3.1) we get, for the reduced path integral, 
$$Z_R = \int d(pq)d(AB)d(c\bar c)e ^{-\int dxdt~\bigl\lbrack p\dot q -
 H(p,q) + A 
\Phi + B\chi + \bar c\lbrace [FP] +
 [BFV]\partial_t\rbrace c\bigr\rbrack }\eqno(3.10)$$
Since, in general, $\chi$ may depend on $ A$ as well as $p$ and $q$, the 
above expression can be used to specialise to either the temporal gauge, for 
which $[FP] = 0$, or to gauges which do not depend on the  
Lagrange multipliers, for which $[BFV] = 0$, with equal facility.   
In the latter case, we can integrate over $ A, B,$  
and the remaining ghosts $\bar c $ and $c$ to obtain the standard 
Faddeev-Popov result [7] {\it viz.} 
$$Z_R = \int d(pq)\delta (\Phi )\delta (\chi ) \mid\mid [FP]\mid\mid 
e^{- \int dxdt~[p\dot q - H(p, q)] } \eqno(3.11)$$
In contrast, for temporal (or ghost-free) gauges, $\chi\equiv A \approx 0$ and 
$\chi\equiv \dot A \approx 0$ we obtain, if we ignore intricacies regarding 
zero modes,   
$$ Z_R = \int d(pq)\mid\mid (\partial_t)\mid \mid 
e^{ -\int dxdt~ [p\dot q - H(p, q)]}\eqno(3.12)$$
Thus, in this case, we obtain the unconstrained phase space path integral 
modified by the determinant for a free field.  
The purpose of the more general formula for the path integral reduction 
factor in Eq. (3.10) is thus clear. It allows us to treat 
the WZW gauge for which 
$A = 0$ on the same footing as other gauges which do not involve the 
Lagrange multipliers. In the following we shall generalise the above results  
to the case at hand {\it viz.} the WZW$\rightarrow $ Liouville reduction.  
\vfil\eject
\centerline {\bf {IV. THE PATH INTEGRAL REDUCTION PROCEDURE}}
\bigskip
Armed with the basic details about the classical WZW $\rightarrow $ Liouville 
reduction and the Batalin-Fradkin-Vilkovisky formalism for quantizing 
constrained systems, from the previous sections, we may now return to the  
problem of constructing the corresponding quantum reduction in terms of 
the phase space path  
integral. However, our application of the BFV formalism to 
the present problem differs from the standard approach reviewed in the last 
section in two respects. First, because we are dealing with independent left 
handed and right handed constraints, it is convenient to replace the standard 
BFV formalism by a light-cone BFV formalism. This is done by replacing 
the space and time directions by the two branches of the light-cone 
parametrised by the light-cone coordinates defined in Eq. (2.2). It is  
important, however, to state that since we use the Euclidean space formulation 
of the path integral, these light-cone coordinates 
actually get converted into holomorphic and anti-holomorphic 
coordinates.  
As a consequence 
of this all the fields in the theory will be functions of the latter  
complex coordinates and any function which depends only on $z_r$ or 
$z_l$ will be a holomorphic or anti-holomorphic function. This fact will 
have important repercussions in the next section. Second    
because the straightforward BFV formalism does not respect conformal 
invariance it has to be modified. We shall modify it in such a way that      
the  conformal invariance is manifest at each stage.   

We begin by noting that the correct phase space path integral measure 
for the unconstrained WZW model is the symplectic  
measure $d(\alpha\beta\gamma
\pi_\alpha\pi_\beta\pi_\gamma )$. This is because an integration over 
the momenta with this measure produces the configuration space path 
integral with the correct group invariant measure $d(e^{-2\beta }\alpha
\beta\gamma )$.  
$$ \eqalign {I_{WZW}(j) &= \int d(\alpha\beta\gamma\pi_\alpha\pi_\beta
\pi_\gamma )~e^{-\int d^2z~[\pi_\alpha\dot\alpha + \pi_\beta\dot\beta +
 \pi_\gamma\dot\gamma -  H_{WZW} + j\beta ]}\cr
&= \int d(e^{-2\beta }\alpha\beta\gamma )  
~e^{-\int d^2z~[L_{WZW} + j\beta ]}}\eqno(4.1)$$  
In the above formula for the Schwinger functional,  
$L_{WZW}$ stands for the Wess-Zumino-Witten Lagrangian    
density and $j$, as usual, stands for an external source. The source is 
attached only to $\beta $ on account of the proposed reduction. 

As discussed in detail in the previous section,
the imposition of the constraints, by means of the BFV formalism, 
will bring into the phase space path    
integral a reduction factor which involves the Lagrange multipliers, the 
ghosts, and their conjugate momenta. In the following we shall proceed to  
construct this factor.  
As a first step towards constructing the reduction factor, we write down the  
expression for the nilpotent BRS charge $\Omega$, following the 
usual BFV prescription namely,  
$$ \Omega \equiv \Omega_r + \{r\leftrightarrow l\},~~~ \Omega_r = \int dz_r~ 
\Omega_r (z)\eqno(4.2) $$ 
where 
$$\Omega_r (z) = c_r(z)\phi_r(z) +  b_r(z)B_r (z)\eqno(4.3) $$ 
In the above expression, $c_r$ and $b_r$ are ghost fields and $B_r$ is the  
momentum conjugate to the Lagrange multiplier field $A_r$ to be introduced  
shortly. The exact splitting of the BRS charge into left and right sectors is 
to be expected because the constraints we are imposing are completely 
independent of each other. For the same reason, the gauge fixing fermion also 
splits into left and right parts. The expression for the right part, 
$\bar \Psi_r$ is given by 
$$\bar \Psi_r (z) = \bar b_r(z)A_r(z) + \bar c_r(z)\chi_r(z)\eqno (4.4)$$ 
A similar expression holds for the left part $\bar \Psi_l$. $\chi_r$ in the 
above equation is the gauge fixing condition for the constraint $\phi_r$. As  
a consequence of the left-right splitting, the reduction factor $F$ factorises  
$$ F = F_r F_l \eqno(4.5)$$ 
$F_r$ and $F_l$ being the corresponding factors for the right and left 
reductions respectively. We shall therefore restrict our attention henceforth 
to one of the sectors. Identical considerations apply naturally to the other 
sector. Notice that there are no higher order terms in the ghosts in the  
expression for the BRS charge. This is  because the constraints have  
exactly vanishing Poisson brackets. It is straightforward to check that  
the Poisson bracket of the improved Hamiltonian density with the BRS  
charge given above is identically zero {\it i.e.} the Hamiltonian 
is gauge invariant. 
The reduction factor $F_r$ can be written as 
$$F_r = \int d(B_rA_r\bar c_rc_r\bar b_rb_r) e^{-\int d^2z~
[\bar b_r\partial_lc_r + \{\Omega_r, \bar\Psi_r\}]}\eqno(4.6)$$
This factor differs from the standard BFV one only in the replacement of 
$\dot c_r$ by $\partial_lc_r$ due to the fact that $\partial_l$ and 
$\partial_r$ play the  
role  of the time derivative in the right-hand and left-hand sectors 
respectively. 

As mentioned earlier, the straightforward application of the above BFV
formalism is not expected to respect  
conformal invariance. This can be seen as follows. The physical 
(Liouville) gauge is defined by the  
condition $\chi_r \equiv  \alpha \approx 0 $. The important point to 
note is that a derivative of $\alpha$  would not suffice 
to fix the gauge completely. Accordingly, the natural conformal 
weight for $\chi_r$ is  
$$\omega (\chi_r ) = (0, 1)\eqno(4.7)$$  
We shall now show that it is not possible to satisfy this condition 
without making some modifications.  
Since $\Omega_r$ generates gauge transformations, it is  
required to be a conformal scalar. And since the Action is a scalar, 
the gauge fixing fermion $\bar \Psi_r$ is required to have a  
conformal weight (1, 1). Using the 
fact that the constraint $\phi_r$ is a conformal scalar, the above 
two requirements translate into the following equations for the weights 
of the various fields respectively  
$$ \omega (c_r) = \omega (b_r) + \omega (B_r) = (0, 1) \eqno(4.8a)$$ 
and 
$$\omega (\bar c_r) + \omega (\chi_r) =
 \omega (\bar b_r) + \omega (A_r) = (1, 1) \eqno(4.8b)$$ 
On the other hand, the conventional Poisson brackets for the fields
$$\{A_r(z), B_r(z')\} = \{b_r(z), \bar c_r(z')\} = \{c_r(z), \bar b_r(z')\} 
= \delta (z_r - z_r')\eqno(4.9)$$
imply that the conformal weights for the fields satisfy  the following equations
$$ \omega (A_r) + \omega (B_r) = \omega (b_r) + \omega (\bar c_r) = 
\omega (c_r) + \omega (\bar b_r) = (0, 1) \eqno(4.10)$$
It is easy to see that the set of equations (4.8) and (4.10) are not 
compatible with Eq.(4.7). It is in this sense that the BFV formalism does not 
automatically incorporate conformal invariance. 
 
The way in which we propose to overcome this difficulty is to introduce 
invertible auxiliary fields $e_r$ and $e_l$ with conformal weights 
$$\omega (e_r) = (0, 1),~~~\omega (e_l) = (1, 0)\eqno(4.11)$$
At this stage the only purpose of these fields is to incorporate manifest 
conformal invariance but their significance will become clear later. 
We use these auxiliary fields to define new Poisson brackets   
$$ \{A_r(z), B_r(z')\} = \{ b_r (z), \bar c_r (z')\} = e_l\delta (z_r - z_r'), 
~~~ \{c_r(z), \bar b_r(z') \} = \delta (z_r - z_r')\eqno(4.12) $$ 
Similar modifications 
apply on the left sector in which we introduce the right partner $e_r$. 
Upon using these new Poisson brackets, the requirement (4.10) is replaced by 
$$\omega (A_r) + \omega (B_r) = 
\omega (b_r) + \omega (\bar c_r) = (1, 1),~~~{\hbox {and}}~~~
\omega (c_r) +  \omega (\bar b_r) = (0, 1)
\eqno(4.13)$$  
It is easy to check that the system of equations (4.8) and (4.13) are 
compatible with Eq. (4.7). There is a certain amount of freedom in assigning 
weights to the fields so as to satisfy these equations but for later 
convenience we choose the following assignment 
$$\vbox{ 
\offinterlineskip
\halign{
\strut\vrule # &\vrule # & \vrule # & \vrule # & \vrule # & \vrule
# & \vrule # & \vrule # \hfil\vrule \cr 
\noalign{\hrule}
~$~~\alpha$~ &~$~~\phi_r$~ & ~$~~A_r$~ & ~$~~B_r$~ &~$~~b_r$~ 
& ~$~~\bar b_r$~ & ~$~~c_r$~ & ~~~$~~\bar c_r$ \cr 
\noalign{\hrule}
~~(0, 1)~& ~~(0, 0)~& ~~(1, 1)~& ~~(0, 0)~& 
~~(0, 1)~& ~~(0, 0)~& ~~(0, 1)~ & ~~(1, 0) \cr
\noalign{\hrule}
} }\eqno (4.14)$$ 
The modified reduction factor $F_r$ is defined by  
$$F_r = \int d\Gamma e^{-\int d^2z~[\bar b_r\partial_lc_r +
 \{\Omega_r, \bar \Psi_r\}]}\eqno(4.15) $$
In the above equation we have deliberately refrained from explicitly 
writing down the measure $d\Gamma$ at this stage as it will be constructed 
a little later taking into account the conformal properties of its constituent 
fields. 

In passing, let us also mention that it is easy to verify that with the 
modified ghost algebra, the BRS charge satisfies the nilpotency condition 
$$\{\Omega , \Omega \} = 0 \eqno(4.16a)$$ 
It also  generates the following gauge transformations  
$$\{\Omega_r , \alpha \} = -c_r, ~~~\{\Omega , A_r\} = -e_lb_r\eqno(4.16b)$$
$$\{\Omega , {\bar b}_r \} = \phi_r,~~~ \{\Omega , \bar c_r\}
= B_r e_l\eqno(4.16c)$$
the rest of the brackets being zero. The consistency with respect to the
conformal dimensionality of the above relations is easily verified. Since the   
right hand sides of the ghost Poisson brackets now involve $e_r$ and $e_l$  
which could, in principle, depend on the background field $\beta $, 
the generalised Jacobi identity involving  
$\pi_\beta$ and the two ghost fields $b, \bar c$, or the Lagrange multipliers  
$A, B$ impels $\pi_\beta $ to have non-vanishing Poisson brackets 
with either the set $(b, B)$ or $(\bar c, A)$. We choose the latter 
option as it automatically ensures that the above modifications in the 
algebra of the ghosts do not tamper with the gauge invariance of the 
improved Hamiltonian density. This therefore reconciles the requirements 
of conformal invariance with the standard ingredients of the BFV 
procedure in a consistent manner. 

We may now readily evaluate the all important $\{\Omega_r, \bar\Psi_r\}$ term 
in $F_r$ using the modified algebra for the ghosts given in (4.12), to find    
$$\{\Omega_r , {\bar \Psi_r}\} = -e_l{\bar b_r}b_r + {\bar c}_r [FP]_rc_r +  
{\bar c}_r[BFV]_rb_r + e_lB_r\chi_r + A_r\phi_r\eqno(4.17)$$ 
where $[FP]_r$ and $[BFV]_r$ are conformal scalars defined by 
$$ \{\phi_r (z), \chi_r (z')\} = [FP]_r\delta (z_r -  z_r'),~~~ 
\{ B_r(z), \chi_r (z')\} = [BFV]_r\delta (z_r - z_r')\eqno(4.18)$$
As in section three, we now wish to perform the integration over  
the $\bar b b$ ghosts. Before we carry out these integrations, however,  
we have to define the correct phase space path  
integral measure $d\Gamma$ for the Lagrange multipliers and their conjugate  
momenta as well as for all the ghosts. This is easily done  
from first principles.  

Let $\phi (z)$ be a quasi-primary field with a conformal dimension   
$s  = s_l +  s_r$, $s_l$ and  $s_r$ being the conformal weights corresponding 
to the left and right Virasoros respectively.  
On an arbitrary manifold, we can expand the field as follows  
$$\phi (z) = \sum c_n\phi_n (z) \eqno(4.19)$$
where $c_n$ are constants and $\{\phi_n (z)\}$ constitute a complete set of 
orthonormal functions. The orthonormality condition is expressed in a 
coordinate invariant way through the equation 
$$(\phi_n, \phi_m) = \int dz_r\int dz_l~(e_re_l)e_r^{-2s_r}e_l^{ - 2s_l}
\phi_n^*\phi_m \eqno(4.20)$$  
Thus the correct fields which have the square integrability property in the 
usual sense are scaled by factors of 
$e_r^{{1\over 2} - s_r}e_l^{{1\over 2} -s_l}$.  
Accordingly, the correct functional measure for the fields is 
$d[e_r^{{1\over 2}- s_r}e_l^{{1\over 2} - s_l}\phi ]$. 
Thus fields which have a conformal weight (0, 1) need a factor of 
$({e_l\over e_r})^{1\over 2}$, fields which have a conformal weight (1, 0)  
need  a factor of $({e_r\over e_l})^{1\over 2}$, conformal scalars 
require a factor of $(e_re_l)^{1\over 2}$, and fields which have a 
conformal weight (1, 1) require a factor of $(e_re_l)^{-{1\over 2}}$.     

Such being the general rule for constructing the conformally invariant
measure, we are now in a position to write down the correct expression for
$d\Gamma$. Taking into consideration the assignment of the weights in 
Eq. (4.14), we see that most of the contributions coming from the various 
fields cancel requiring us  to modify the standard  measure 
 $d(B_rA_r\bar b_r b_r\bar c_r c_r)$ by  just  
a factor of $e_l$.  Thus we have for the reduction factor   
$$ F_r = \int d(e_l B_rA_r\bar b_r b_r \bar c_r c_r )
e^{-\int d^2z~ \bigl\lbrack \bar b_r\partial_lc_r - e_l\bar b_rb_r
+ \bar c_r [FP]_r c_r + \bar c_r [BFV]_r b_r + e_lB_r\chi_r
+ A_r\phi_r\bigr\rbrack  }$$
Integrating over the $\bar b b$ fields now gives
$$ F_r  = \int d(B_rA_r\bar c_rc_r) e^{-\int d^2z~~ 
\bigl\lbrack \bar c_r[FP]_r c_r + \bar c_r[BFV]_re_l^{-1}\partial_lc_r 
+ e_lB_r\chi_r + A_r\phi_r \bigr\rbrack }\eqno(4.21)$$
All the results we have obtained above are equally valid in the left    
sector of the reduction and can be obtained simply by exchanging the suffixes  
$r$ and $l$ and interchanging the two entries corresponding to the 
left and right Virasoros in the conformal weights of the fields. We therefore   
have for $F_l$                                          
$$ F_l = \int d(B_lA_l\bar c_lc_l) e^{-\int d^2z~~
\bigl\lbrack \bar c_l[FP]_l c_l + \bar c_l[BFV]_l e_r^{-1}\partial_rc_l 
+ e_rB_l\chi_l + A_l\phi_l\big\rbrack }\eqno(4.22)$$
The full reduction factor that needs to be introduced into the WZW path 
integral is therefore, 
$$ F = \int d(B_rB_lA_rA_l\bar c_rc_r\bar c_lc_l )e^{-\int d^2z
\bigl\lbrack 
\bar c_r[FP]_rc_r + \bar c_r[BFV]_re_l^{-1}\partial_lc_r + e_lB_r\chi_r 
+ A_r\phi_r + \{r\leftrightarrow l\}\bigr\rbrack }\eqno(4.23)$$  
This expression is the conformally invariant generalisation of the
standard BFV reduction factor in Eq. (3.9) for the present theory. Since 
this generalisation introduces non-trivial modifications to the standard 
BFV formalism, we have to  prove that these modifications are indeed 
consistent. We do this by proving an analogue of the Fradkin-Vilkovisky 
theorem for the gauge independence of the path integral of the reduced 
theory within our generalised formalism.  
\bigskip
\bigskip
\noindent ${\underline {\bf Theorem}}$: If the reduction factor $F(\bar \Psi )$ 
is defined as  
in Eqn.(4.23), and the gauge fixing functions $\chi_r$ and $\chi_l$ are 
independent of the fields $B_r$ and $B_l$, as is usually the case,
 the reduced path integral 
$$ I_{\bar \Psi}(j) = \int d(\alpha\beta\gamma\pi_\alpha\pi_\beta\pi_\gamma )
e^{-\int d^2z~[\pi_\alpha\dot\alpha + \pi_\beta\dot\beta + \pi_\gamma\dot\gamma 
- H_{WZW} + j\beta ]} \times F(\bar \Psi )\eqno(4.24)$$ 
is independent of $\bar \Psi$. In fact, 
$$ I_{\bar\Psi} (j) = \int d(e_r^{-1}e_l^{-1}\beta ) e^{-\int d^2z 
[(\partial_\mu\beta )(\partial^\mu\beta ) + m_rm_le^{2\beta } + j\beta ]}
 \eqno(4.25)$$
which is manifestly independent of $\bar\Psi$.  
\bigskip
\noindent ${\underline {\bf Proof}}$: Since the gauge fixing 
functions $\chi_r, \chi_l$ are independent of $B_r$ and $B_l$, 
we may integrate over the $B$
 fields in Eq. (4.23) to get 
$$F = \int d(e_r^{-1}e_l^{-1}A_rA_l\bar c_r c_r\bar c_l c_l ) 
\delta (\chi_r)\delta (\chi_l)
e^{-\int d^2z \bigl\lbrack\bar c_r[FP]_rc_r +
 \bar c_r [BFV]_re_l^{-1}\partial_lc_r 
+ A_r\phi_r + \{r\leftrightarrow l\}\bigr\rbrack }\eqno(4.26)$$  
Using the fact that the 
constraints $\phi_r$ and $\phi_l$ are expressible in terms of the momenta 
$\pi_\alpha$ and $\pi_\gamma$ through Eq. (2.14b), we can introduce the above 
reduction factor into Eq. (4.24) and integrate over 
the momenta $\pi_\alpha , \pi_\beta , \pi_\gamma $ to get the gauged  WZW    
model 
$$I_{\bar\Psi}(j)  = \int d(e_r^{-1}e_l^{-1}\alpha\beta\gamma 
A_rA_le^{-2\beta})\delta(\chi_r) 
\delta (\chi_l)~e^{-\int d^2z~[L_{GWZW} + j\beta ]} \times G \eqno(4.27a)$$ 
where $L_{GWZW}$ stands for the Lagrangian density of the gauged
Wess-Zumino-Witten model and is given by  
$$L_{GWZW} = (\partial_\mu\beta )(\partial^\mu\beta ) + e^{-2\beta}
(\partial_r\gamma + A_l)(\partial_l\alpha + A_r) -A_lm_l - A_rm_r
\eqno(4.27b)$$ 
and  
$$\eqalign {G &= \int d(\bar c_rc_r\bar c_lc_l)e^{-\int d^2z~\bigl\lbrack\bar
 c_r[FP]_rc_r +  
\bar c_r[BFV]_re_l^{-1}\partial_lc_r + \{l\leftrightarrow r\}\bigr\rbrack }\cr 
&= \int d(\bar c_rc_r\bar c_lc_l)e^{-\int d^2z~\bigl\lbrack\bar c_r
[{\partial\chi_r\over\partial\alpha} - {\partial\chi_r\over\partial  A_r}
\partial_l]c_r + \{r\leftrightarrow l\} \bigr\rbrack }}
\eqno(4.27c)$$ 
stands for the ghost-factor. Notice that this expression differs from 
what one might naively expect for the gauged WZW path integral because of the 
appearance of the auxiliary fields $e_r$ and $e_l$ in the measure. 
But, as is amply 
clear from the foregoing, 
these are precisely the factors that enable us to carry out the reduction 
in a conformally invariant fashion. We now define the shifted fields  
$$A_l \rightarrow \bar A_l = A_l + \partial_r\gamma ,~~~A_r
\rightarrow \bar A_r = A_r + \partial_l\alpha \eqno(4.28)$$ 
Notice that $\partial_l\alpha$ and $\partial_r\gamma$ can always be absorbed 
by a redefinition of the $A$ fields as above, although the presence of
zero modes may not always  
allow us to completely   
eliminate the $A$ fields themselves by shifting 
$\alpha$ and $\gamma$ appropriately.    
In terms of the shifted fields Eqs. (4.27) become 
$$I_{\bar\Psi}(j) = \int d(e_r^{-1}e_l^{-1}\alpha\beta\gamma\bar A_r\bar A_l
e^{-2\beta })\delta 
(\chi_r)\delta (\chi_l)e^{-\int d^2z~[L_{GWZW} + j\beta ]}\times G\eqno(4.29a)$$
and 
$$L_{GWZW} =  (\partial_\mu\beta )(\partial^\mu\beta ) 
+ e^{-2\beta }\bar A_r \bar A_l - \bar A_lm_l - 
\bar A_rm_r  \eqno(4.29b) $$
respectively, where we have dropped total derivative terms that appear
 in shifting the 
$m_r$ and $m_l$ dependent terms. 
The ghost factor $G$ has the following nice interpretation in terms of the 
shifted fields. Recall that the gauge fixing condition $\chi_r$ is, in 
general, a function of both $\alpha$ and the Lagrange multiplier $A_r$ 
which are independent of each other. If we work in terms of the shifted 
fields defined above, this is no longer true and we have
$$ \bigl\lbrack {\partial\chi_r\over\partial\alpha }\bigr\rbrack_{\bar A_r} 
 = \bigl\lbrack {\partial\chi_r\over \partial\alpha }\bigr\rbrack_{A_r}  
- \bigl\lbrack{\partial\chi_r\over \partial A_r}\bigr\rbrack_\alpha 
\partial_l \eqno (4.30)$$ 
where the partial derivatives in the above equation are to be taken  
keeping the fields appearing as subscripts fixed.  
Notice that the right hand side of the above equation is just the argument 
in the determinant 
that results from performing the ghost integrations in Eq. (4.27c). Taking   
this into account, the measure in Eq. (4.29a) becomes 
$$\eqalign { d(e_r^{-1}e_l^{-1}\alpha\beta\gamma\bar A_r\bar A_le^{-2\beta})
 \delta (\chi_r)
\delta (\chi_l) \mid\mid&\bigl\lbrack{\partial\chi_r\over\partial\alpha}\bigr
\rbrack_{\bar A_r}
\bigl\lbrack {\partial\chi_l\over\partial\gamma }\bigr\rbrack_{\bar A_l}   
\mid\mid\cr = d(e_r^{-1}e_l^{-1}\alpha\beta\gamma\bar A_r\bar A_le^{-2\beta})
\delta (\alpha )
\delta (\gamma )} \eqno(4.31) $$  
The $\alpha$ and $\gamma$ integrations now drop out to yield   
$$ I_{\bar\Psi}(j) = \int d(e_r^{-1}e_l^{-1}\beta \bar A_r\bar A_le^{-2\beta})~
e^{-\int d^2z~L_{GWZW}}
 \eqno(4.32a)$$ 
Carrying out the gaussian integration over the $\bar A$ fields we then 
obtain 
$$ I_{\bar \Psi}(j) = \int d(e_r^{-1}e_l^{-1}\beta ) e^{-\int d^2z~  
[ (\partial_\mu\beta )(\partial^\mu\beta ) - m_rm_le^{2\beta} + j\beta ]}  
\eqno(4.32b)$$
as required. We have therefore proved that the Fradkin-Vilkovisky theorem 
can be generalized to include conformal invariance.  
Although this was done within the context of the 
WZW $\rightarrow$ Liouville reduction, it is clear that the principle 
is sufficiently general to transcend the domains of the present theory 
and should apply to all conformally invariant gauge theories.  

We will now discuss the role played by the auxiliary fields.
 The crucial point to note  is that they appear in the final result and 
because they have non-zero conformal weights, they can not be set equal 
to unity without breaking conformal invariance. Thus they are an intrinsic 
part of the reduction.   

On the other hand they appear only in the  measure and only  
in the form of the  product $e_re_l$  
which has a conformal weight (1,1). It is this fact that allows us to  
use them without introducing any new dynamics since the conformal weights 
allow us to make the following natural  
identification   
$$e_re_l \equiv e^{2\beta }\eqno(4.33)$$  
Moreover, if we regard  $e^{2\beta}$ as  $\sqrt g$, where $g$ is the  
determinant of a two-dimensional metric, 
Eq.(4.33) allows us to immediately recognise the $e_r$ and $e_l$ fields as
the two components of a  {\it Zweibein}.    
It is interesting to note that had the reduction not been 
left-right symmetric, other combinations of the components of the 
{\it Zweibein} would have occured in the final results and these 
would have corresponded to genuine external fields. 
Using Eq. (4.33) in Eq. (4.32b), and reintroducing the WZW coupling constant 
$\kappa $ we get 
$$I(j) = \int d(e^{-2\beta }\beta )e^{-{\kappa\over \pi}
\int d^2z~[(\partial_\mu\beta )
(\partial^\mu\beta ) - m_rm_le^{2\beta } + j\beta  ]}\eqno(4.34)$$  
where the allusion to $\bar \Psi$ has been dropped for obvious reasons. 

As is well known [8], the exponential factor in the measure of Eq.(4.34)
corresponds to the conformal anomaly and  can  
be removed by making a suitable shift in the WZW coupling constant to yield,   
$$I (j) = \int d\beta e^{-{(\kappa -2)\over \pi}\int d^2z ~
[\partial_\mu\beta\partial^\mu
\beta - m_rm_le^{2\beta} +j\beta  ]}\eqno(4.35)$$ 
This then is the Liouville theory that is 
the result of the reduction. It is well-known [9] that the Virasoro centre 
for the above theory has the form   
$$c=\hbar + 6\bigl\lbrack \sqrt{k-2\hbar} + {\hbar\over\sqrt{k-2\hbar}}
\bigr\rbrack^2\eqno(4.36)$$   
where $k = {\kappa \over 2\pi}$. In the next section, we shall give a simple 
interpretation of this formula.  
\bigskip 
\centerline {\bf {V. COMPARISON WITH EARLIER PATH-INTEGRAL RESULTS }}
\bigskip
Although most quantized treatments of WZW $\rightarrow$ Liouville 
reductions use the canonical formalism [2], the functional integral formalism  
was considered in references [4, 10]. In these references the Faddeev-Popov  
method was used and led to a Liouville theory.  
These references study the  
path integral in two special gauges, namely the Liouville gauge  
and the WZW gauge. It would therefore be reassuring 
to redo our analysis in these gauges in order to compare our 
results with these earlier works. In fact these gauges highlight the 
roles of the anomaly and the zero modes respectively. 
We first examine the Liouville gauge.  
\bigskip  
\bigskip 
\noindent ${\underline {\bf Liouville~ Gauge}}$: In this gauge we have 
$$\chi_r \equiv \alpha \approx 0,~~~~\chi_l\equiv\gamma\approx 0\eqno(5.1)$$
and hence it follows from Eqs. (4.18) that,  
$$ [FP]_{r, l} = -1,~~~ {\hbox {and}}~~~ [BFV]_{r, l} = 0\eqno(5.2)$$
Substituting the above equalities into our expression for the generalised 
reduction factor Eq. (4.26), we get 
$$ F = \int d(e_r^{-1}e_l^{-1}A_rA_l\bar c_rc_r\bar c_lc_l)\delta (\alpha ) 
\delta (\gamma )e^{-\int d^2z~[\bar c_rc_r + A_r\phi_r + \{r\leftrightarrow 
l\}]}\eqno(5.3)$$ 
Doing the $A$ integrations and the trivial ghost integrations we find 
that the BFV reduction factor is just   
$$ F = det \mid\mid e_r^{-1}e_l^{-1}\mid\mid \delta (\phi_r )\delta (\phi_l)
\delta (\alpha )\delta (\gamma ) \eqno(5.4)$$ 
Inserting this factor into the unconstrained WZW phase space path integral and  
carrying out the various delta function integrations as well as the 
gaussian $\pi_\beta$
integration we get, as expected, Eq. (4.32b). This result differs from the 
result of earlier path integral formulations of the problem by the appearance 
of the factor $ \mid\mid e_r^{-1}e_l^{-1}\mid\mid $ in the measure.  
Since $(e_re_l)^{-1} = e^{-2\beta}$ according to Eq. (4.33), the use of 
the {\it Zweibein} changes the Liouville measure from $d(\beta ) $ to 
$d(e^{-2\beta }\beta )$ and thus produces a conformal anomaly.  
As already mentioned, the insertion of this factor   
in the measure is equivalent 
to a change of $k$ to $k-2$ in the exponent and thus leads to a change  
$$\hbar +6\Bigl(\sqrt{k}
+{\hbar \over \sqrt{k}}\Bigr)^2\qquad \rightarrow \qquad 
\hbar +6\Bigl(\sqrt{k -2\hbar}
+{\hbar \over \sqrt{k - 2\hbar}}\Bigr)^2 \eqno (5.5)$$
in the Virasoro centre. The difference between the two expressions 
in Eq. (5.5) is the discrepancy that was 
mentioned in the Introduction and can now be seen to be due to the fact 
that 
the {\it Zweibein} was not used in the earlier papers.  
\bigskip
\bigskip 
\noindent ${\underline {\bf WZW~ Gauge}}$: This gauge is the analogue of 
the temporal gauge in QED and is  
defined by setting the Lagrange multipliers equal to zero, modulo zero  
modes.  On a compact 2-space (whose compactness, we recall, is necessitated 
by the Liouville potential, which in turn is present because of the non-zero 
constants $m_r$ and $m_l$) there is just one zero-mode for each $A$.  
To see this let us consider $A_r$, for example, and decompose it 
according to  
$$A_r = A_r^0 + \hat A_r,~~~\hat A_r = \partial_l\lambda_r,~~~
\int d^2z~ (e_re_l)^{-1}A_r^0\hat A_r = 0 \eqno(5.6)$$
{\it i.e.} into a part $\hat A_r$ that can be gauged away and its orthogonal 
complement $A_r^0$. In the above equation the gauge transformation parameter 
$\lambda_r$ has a conformal weight $\omega (\lambda_r ) = (0, 1)$.
The factor $(e_re_l)^{-1}$ in the integral comes from
the requirement that the orthogonality condition be conformally invariant. 
Since the orthogonality must hold for arbitrary $\lambda_r$, 
it follows from a simple partial integration that 
$$\partial_l (e_r^{-1}e_l^{-1}A_r^0) = 0~~~ {\hbox {or}}~~~ 
A_r^0 = e_re_lf(z_r)\eqno(5.7)$$
where $f(z_r)$ is an arbitrary holomorphic function. However, since there are  
no holomorphic functions on a compact Riemann surface except the constant 
functions [11], we see that $f(z_r)$ must be constant and thus the only 
normalised zero-mode is  
$$A^0_r = {e_re_l\over \sqrt V}~~~{\hbox {where}}~~~
V = \int d^2z~e_re_l = \int d^2z~e^{2\beta }\eqno(5.8)$$ 
A similar expression holds for $A_l^0$. Thus the WZW gauge is 
$$\chi_r\equiv e_l^{-1}\hat A_r\approx 0,~~A_r^0 = \mu_r{e_re_l\over\sqrt V}
~~~{\hbox {and }}~~~\chi_l\equiv e_r^{-1}\hat A_l\approx 0,~~
A_l^0 = \mu_l{e_re_l\over \sqrt V} \eqno(5.9) $$ 
where the $\mu$'s  are arbitrary constants. Notice that this is a complete 
gauge fixing because it determines the gauge parameter $\lambda_r$ up
to a function $\lambda (z_r)$
and the only such function is a constant which must be zero because 
$\lambda_r$ has a conformal weight (0, 1). Similar considerations apply for 
$\lambda_l$.  
The measure for the Lagrange multipliers now becomes 
$$ d(e_r^{-1}e_l^{-1}A_rA_l) = d(\mu_r\mu_l)
d(e_r^{-1}e_l^{-1}\hat A_r\hat A_l) \eqno(5.10)$$  
The expressions for the $\chi$'s imply that 
$$[FP]_{r, l} = 0, ~~~[BFV]_{r, l} = -1\eqno(5.11)$$
Substituting the above results in Eq. (4.27) and doing the ghost 
integrations yields  
$$\eqalign {I(j) = \int d(e_r^{-1}e_l^{-1}\alpha\beta\gamma \mu_r\mu_l\hat A_r
\hat A_l e^{-2\beta })\delta (e_l^{-1}\hat A_r)\delta (e_r^{-1}\hat A_l)
&\mid\mid e_l^{-1}\partial_l e_r^{-1} \partial_r \mid\mid \cr 
&\times e^{-\int d^2z~ [L_{GWZW} + j\beta ]}} \eqno(5.12)$$
The integration over the $\hat A$ fields can now be performed using the 
gauge fixing delta functions to yield 
$$I(j) = \int d( \alpha\beta\gamma e^{-2\beta }\mu_r\mu_l ) 
\mid\mid e_l^{-1}\partial_le_r^{-1}\partial_r \mid\mid  
~e^{-\int d^2z~[L^0_{GWZW} + j\beta ]}\eqno(5.13a)$$ 
where 
$$\eqalign {L^0_{GWZW} &= (\partial_\mu\beta )(\partial^\mu\beta ) + 
e^{-2\beta }
(\partial_l\alpha + A_r^0 )(\partial_r\gamma + A_l^0 ) - m_rA_r^0 
- m_lA_l^0 \cr 
&= (\partial_\mu\beta )(\partial^\mu\beta ) + e^{-2\beta }(\partial_l\alpha )
(\partial_r\gamma )  + \mu_r\mu_l  
- {\mu_r m_r\over\sqrt V}e_re_l  
- {\mu_l m_l\over\sqrt V}e_re_l  
} \eqno(5.13b) $$
In arriving at the above equation we have used the expressions for $A^0$'s 
in Eq. (5.9) and the 
equality $e_re_l = e^{2\beta }$ in Eq. (4.33). 
The integration over the $\alpha$ and $\gamma $ fields produces a factor that  
exactly cancels the $e^{-2\beta}$ factor in the measure and the
$\mid\mid\partial_r\partial_l\mid\mid$ factor in the fermionic determinant. 
The path integral therefore reduces to 
$$I(j) = \int d(e_r^{-1}e_l^{-1}\beta \mu_r\mu_l )
e^{-\int d^2z ~[(\partial_\mu\beta )(\partial^\mu\beta ) + j\beta 
+ e_re_l({\mu_r\mu_l\over V} - {\mu_l\over \sqrt V}m_r 
-{\mu_r\over \sqrt V}m_l )]}\eqno(5.14)$$
The zero modes can now be integrated without further ado to produce  
$$ I(j) = \int d(e_r^{-1}e_l^{-1}\beta ) e^{-\int d^2z~[(\partial_\mu\beta )
(\partial^\mu\beta ) - m_rm_le^{2\beta } + j\beta ]}\eqno(5.15)$$
where we have used Eq. (4.33) to set $e_re_l = e^{2\beta }$.     
We have thus verified that the WZW gauge produces the same 
result as the Liouville gauge. From the foregoing discussion it is 
clear that the Liouville potential is actually due to the  
zero-modes.   

However, our interest here is not in verifying that the WZW gauge leads 
to the correct result but in comparing the final results with the 
expressions obtained in the previous papers [4, 10]. In those papers  
the zero-modes were neglected and the WZW gauge was defined as 
 $A_r = A_l\approx 0$. As a result, the final expression in the WZW gauge 
was the same as in Eq. (5.15) but without the Liouville potential. 
The omission of the Liouville potential  actually 
made no difference to the final results because the purpose of those papers 
was to compute the Virasoro centre; and for that purpose the only  
role of the Liouville potential is to require the use of the  
improved Virasoro operators. Since the impoved Virasoros were used in any 
case, the result obtained for the centre in those papers was the correct 
one.  

The fact that the earlier computations of the Virasoro centre in the WZW gauge  
are still valid allows us to draw two interesting conclusions. First, 
since the expression for the Virasoro centre is independent of $m_r$ and 
$m_l$, there is a smooth transition for the Virasoro algebra  
to the case $m_l=m_r=0$ even though the reduced system in the
latter case does not require the 2-space to be compact.  
Second, since the earlier WZW-gauge computations are valid,  
they provide an interesting interpretation of 
the formula for the Virasoro centre in the Liouville theory which is not at 
all obvious in the context of the Liouville theory itself. In fact,  
they show that if the Liouville theory formula for the centre is expanded 
according to 
$$C=\hbar +6\Bigl(\sqrt{k-2\hbar}
+{\hbar \over \sqrt{k-2\hbar}}\Bigr)^2=
{3k\hbar \over k-2\hbar} -2\hbar  + 6k \eqno(5.16) $$
it is just the sum of three independent centres namely, 
the centre for the $SL(2,R)$ WZW 
model, the centre for the ghosts, and the centre for the classical improvement
term. The results of [4, 10] show that a similar 
interpretation exists for Toda theories. 
\bigskip
\centerline {\bf {VI. SUMMARY AND CONCLUSIONS}}
\bigskip
We have introduced a generalization of the Batalin-Fradkin-Vilkovisky 
formalism which allows us to incorporate conformal invariance into the usual  
procedure for the path integral quantization of systems with first-class 
constraints. Although we have done 
this only for WZW $\rightarrow$ Liouville reduction in two dimensions it is  
clear that the procedure should apply to all conformally invariant reductions 
and should be independent of the dimension. In later papers we hope 
to apply it to WZW $\rightarrow$ Toda and Goddard-Olive reductions the latter  
of which will require a further generalization of our analysis to include 
second class constraints. 
An essential feature of our procedure is the introduction of a {\it Zweibein} 
which makes the conformal invariance manifest at each stage of the reduction.  
The two components of this {\it Zweibein} appear in the final theory only as 
products of the form $e_re_l = e^{2\beta }$ where $\beta$ is the  
Liouville field, and thus introduce non-trivial modifications of the 
reduction (actually a conformal anomaly) without introducing new fields. 
Our main result is that, in spite  
of the conformal anomaly, an analogue of the Fradkin-Vilkovisky  
theorem is still valid. 

An interesting feature of the WZW $\rightarrow$ Liouville 
(or indeed Toda) reductions is that the first-class constraints are 
obtained by setting the momenta not equal to zero but to constants $m_l$ 
and $m_r$. When these constants are not zero the gauge fields 
(Lagrange multipliers)  
have zero-modes and    
it is precisely these zero-modes that produce the exponential 
Liouville interaction.  

Earlier papers, in which the straightforward Faddeev-Popov  
formalism was used, did not produce the conformal anomaly in the 
Liouville gauge, which led to a   
discrepancy in the expression for the Virasoro centre in the Liouville 
and WZW gauges.  
Our analysis traces the origin of this  discrepancy to the fact that 
the standard Faddeev-Popov formalism, in spite of its appearance, is not  
conformally invariant. A modification using a {\it Zweibein}  produces  
a formalism which is both gauge and conformal invariant. 
\bigskip
\bigskip
We would like thank L. Feh\'er, M. Fry, I. Sachs, and 
I. Tsutsui for interesting discussions. 
\vfil\eject
\centerline {\bf {REFERENCES}}
\bigskip
\item {1. } F. A. Bais, T. Tjin and P. Van Driel, Nuc. Phus. {B 357} (1991) 
632; V. A. Fateev and S. L. Lukyanov, Int. J. Mod. Phys. {A3} (1988) 507;
S. L. Lukyanov, Funct. Anal. Appl. {22} (1989) 255; P. Forg\'acs, A. Wipf,
J. Balog, L. Feh\'er, and L. O'Raifeartaigh, Phys. Lett. {B227} (1989)214;
J. Balog, L. Feh\'er, L. O'Raifeartaigh, P. Forg\'acs and A. Wipf, Ann. Phys. 
{203} (1990) 76; Phys. Lett {B244}(1990)435; L. Feh\'er, 
L. O'Raifaertaigh, P. Ruelle, I. Tsutsui and A. Wipf, Phys. Rep. { 222}
No. 1, (1992)1; P. Bouwknegt and K. Schoutens Eds. W Symmetry (Advanced 
series in mathematical physics, 22) World Scientific, Singapore (1995).  

\item {2. } A. Bilal and J. L. Gervais, Phys. Lett {B206} (1988) 412;
Nuc. Phys. { B314} (1989) 646; { B318} (1989) 579; O. Babelon, Phys. 
Lett. { B215} (1988) 523, T. Hollowod and P. Mansfield, Nuc. Phys. 
{ B330} (1990) 720; P. Mansfield and B. Spence, Nuc. Phys. { B362}  
(1991) 294; P. Bowcock and G. M. T. Watts, Nuc. Phys. { B379} (1992) 63;
C. Ford and L. O'Raifeartaigh, Nuc. Phys. { B460} (1996) 203. 

\item {3. } I. A. Batalin and E. S. Fradkin, in: Group Theoretical Methods 
in Physics, Vol II (Moscow, 1980); I. A. Batalin and G. Vilkovisky, 
Phys. Lett. { B69} (1977) 309; E. S. Fradkin and G. A. Vilkovisky, 
Phys. Lett. { B55} (1975) 224; M. Henneaux, Phys. Rep { 126} No.1, 
(1985) 1. 

\item {4. } L. O'Raifeartaigh in 1992 CAP/NSERC Summer Institute in 
Theoretical Physics, Quantum Groups, Integrable Models and Statistical 
Systems Eds. J. LeTourneux and Luc Vinet (World Scientific 1993). 

\item {5. } E. Witten, Comm. Math. Phys. { 92} (1984) 483; P. Goddard and 
D. Olive, Int. J. Mod. Phys. { A1} (1986) 303; P. Di Francesco, P. Mathieu,
and D. S\'en\'echal, Conformal Field Theory, Graduate Texts in Contemporary 
Physics, (Springer-Verlag New York, Inc. 1997).     

\item {6. } I. Tsutsui and L. Feh\'er, Prog. Theor. Phys. Suppl. { 118} 
(1995), 173. 

\item {7. } L. D. Faddeev, Theor. Math. Phys. { 1} (1970) 1;
 L. D.  Faddeev and  A. A. Slavnov, Gauge Fields, introduction 
to quantum theory. (Benjamin Cummins, Reading, Massachusetts, 1980).  

\item {8. } N. D. Birrell, P. C. W. Davies, Quantum Fields in Curved Space, 
(Cambridge University Press 1982); P. Di Francesco {\it at al} ref. 5. 

\item {9. } T. L. Curtright and C. B. Thorn, Phys. Rev. Lett. { 48} 
(1982) 1309, E. D' Hoker and R. Jackiw, Phys. Rev. { D26} (1982) 3517; 
Phys. Rev. Lett. { 50} (1983) 1719.  

\item {10. } L. O'Raifeartaigh, P. Ruelle and I. Tsutsui, Phys. Lett. 
{ B258} (1991) 359. 

\item {11. } Phillip A. Griffiths, Introduction to Algebraic Curves, 
Vol. { 76} Translations of Mathematical Monographs, American 
Mathematical Society (1989). 
\vfil\eject\end